\documentclass{jpsj3}
\usepackage{txfonts}

\usepackage[dvipdfmx]{graphicx}
\usepackage{bm}

\newcommand{\be}{\begin{equation}}
\newcommand{\ee}{\end{equation}}
\newcommand{\eq}[1]{Eq.~(\ref{#1})}
\newcommand{\fig}[1]{Fig.~\ref{#1}}
\def\bea{\begin{eqnarray}}
\def\eea{\end{eqnarray}}

\def\bra{\langle}
\def\ket{\rangle}
\def\vq{{\bf q}}

\def\vk{{\bf k}}

\title{Theoretical insights into electronic nematic order, bond-charge orders, and plasmons in cuprate superconductors} 

\author{Hiroyuki Yamase$^{1,2}$\thanks{yamase.hiroyuki@nims.go.jp}}
\inst{$^1$International Center of Materials Nanoarchitectonics, 
National Institute for Materials Science, Tsukuba 305-0047, Japan \\ 
$^2$Department of Condensed Matter Physics, Graduate School of Science, Hokkaido University, Sapporo 060-0810, Japan}

\recdate{March 31, 2021}

\abst{
The parent compound of high-$T_c$ cuprate superconductors is a Mott insulator described by 
the Heisenberg spin-spin interaction on a square lattice. 
With  carrier doping, the charge degree of freedom becomes active and both spin and charge 
couple to each other, leading to very rich physics including high-$T_c$ superconductivity. 
In this article, we focus on the charge degree of freedom and review theoretical insights 
into the electronic nematic order, bond-charge orders, and plasmons. 
The low-energy {\it charge} dynamics is controlled by the {\it spin-spin} interaction $J$, 
which generates various bond-charge ordering tendencies  including the electronic nematic order. 
The nematic order is driven by a $d$-wave Pomeranchuk instability and is pronounced 
in the underdoped region as well as around van Hove filling in the hole-doped case; 
the nematic tendency is weak in the electron-doped region. 
Nematicity consistent with the $d$-wave Pomeranchuk instability was reported for hole-doped cuprates in 
various experiments such as inelastic neutron scattering, angle-resolved photoemission spectroscopy, Compton scattering, 
electronic Raman scattering, and  measurements of Nernst coefficients and magnetic torque. 
Although the $t$-$J$ and Hubbard models correctly predicted the proximity to 
the nematic instability in cuprates far before the experimental indications were obtained, 
full understanding of the charge ordering tendencies in hole-doped cuprates still requires further theoretical studies. 
In electron-doped cuprates, on the other hand, 
the $d$-wave bond-charge excitations around momentum $\vq\approx (0.5\pi,0)$ explain 
the resonant x-ray scattering data very well.  
Plasmon excitations are also present and 
the agreement between the large-$N$ theory of the $t$-$J$ model and 
resonant inelastic x-ray scattering measurements is nearly quantitative 
in both hole- and electron-doped cuprates. 
Theoretically the charge dynamics in cuprates is summarized as a dual structure in energy space: 
the low-energy region scaled by $J$, where the nematic and various bond-charge orders are relevant, and the high-energy region 
typically larger than $J$, where plasmons are predominant. 
We hope that the present article serves for a sound basis toward 
further experimental and theoretical studies on the origin of the pseudogap and ultimately the high-$T_c$  mechanism. 
}

\begin{document}
\maketitle

\section{Introduction}
The parent compound of high-temperature cuprate superconductors is 
a charge-transfer type Mott insulator and exhibits the Ne\'el state \cite{imada98}. 
There is a large gap between Cu $3d^9$ and $3d^{10}$ states due to the strong on-site 
Coulomb repulsion and electrons are occupied up to the 3d$^9$ state.  
Inside the gap there are oxygen $2p$ states, which are in the closed shell. 
The gap between Cu $3d^{10}$ state and O $2p$ states is called the charge transfer gap $\Delta_{\rm CT}$, 
which is estimated around 2 eV (Ref.~\citen{uchida91}). 
Below the energy scale of $\Delta_{\rm CT}$, the spin degrees of freedom are only active and 
the system is well described by the Heisenberg spin interaction on a square lattice \cite{chakravarty89}. 

Upon hole doping, holes enter the O $2p$ states and strongly interact with the Cu $3d$ spins to form 
the so-called Zhang-Rice singlets \cite{fczhang88}. The electronic state is then described by the motion 
of the Zhang-Rice singlets in the antiferromagnetic background. This entangled state of the spin and charge 
degrees of freedom is believed to be described by the one-band $t$-$J$ and Hubbard models on a square lattice 
with a large onsite repulsion $U$ (Ref.~\citen{anderson87}). 
On the other hand, electron doping is also possible. Electrons enter the Cu $3d^{10}$ state and no concept of 
the Zhang-Rice singlet formation is necessary. 
Yet,  the essential physics and the theoretical models are 
believed to be the same as the hole-doped case. 
That is, the doped electrons are mobile in the antiferromagnetic background 
with experiencing strong onsite Coulomb repulsion at each Cu site. 

The cuprate physics is nothing less than the physics of a doped Mott insulator on a square lattice, independent of carrier types. 
Furthermore effectively only one band is relevant to the physics related to 
the energy scale below $\Delta_{\rm CT}$. The study of the cuprate physics therefore provides a simple setup to 
explore the very rich physics of a doped  Mott insulator \cite{keimer15}: 
incommensurate spin excitations \cite{thurston89,yamada98}, 
spin resonance \cite{rossat-mignod91}, spin-glass phase \cite{wakimoto00}, 
coupled states of spin and charge degrees of freedom referred to as spin-charge stripes \cite{tranquada95},  
1/8-anomaly \cite{moodenbaugh88,kumagai88,sera89}, charge ordering tendencies which do not seem to couple to 
spins \cite{ghiringhelli12,chang12,achkar12,da-silva-neto15}, 
possible coexistence of superconductivity and magnetism \cite{niedermayer98,kimura99,haug10,mukuda12,kunisada20},   
and needless to say pseudogap \cite{timusk99} and high-temperature superconductivity \cite{bednorz86}. 

While all those phenomena are believed to be ultimately described by the $t$-$J$ and Hubbard models, 
the current situation is still far away from that, because of the difficulty to handle strong electron 
correlation effects in a controllable way and to perform systematic calculations toward the ultimate goal. 
Practically uncontrollable approximations are frequently made and 
various different models other than the $t$-$J$ and Hubbard models are also explored to endeavor to catch 
the essential physics of a doped Mott insulator. 

In this article, we focus on the charge degree of freedom and provide theoretical insights 
into the electronic nematic order (Sec.~2), bond-charge orders (Sec.~3), and plasmons (Sec.~4).  
In Sec.~5, we provide a slightly detailed summary. 
The page limitation did not allow us to cover all interesting works related to the three subjects above. 
In addition, the spin degrees of freedom and a coupling between spin and charge are beyond our scope. 

\section{Electronic nematic physics}
The electronic nematic order breaks the rotational symmetry of the system, 
leaving the other symmetries unbroken. There are two different routes to obtain the nematic order. 

One is to assume the so-called stripe ordered phase, which breaks both rotational and translational symmetries, 
and to envisage that the charge stripes fluctuate due to the low-dimensionality and restore 
the translational symmetry alone \cite{kivelson98}. 
In this scenario, the nematic order is regarded as a vestigial stripe order. 
Consequently the charge stripe order is expected below the nematic phase \cite{misc-FeSCnematic}. 
The so-called charge stripe orders were actually reported in the $t$-$J$ model \cite{white98a,white98b,corboz11,corboz14},  
but such results are in conflict with the exact diagonalization study \cite{hellberg99} 
and the fixed-node Monte Carlo study \cite{hu12} on the same model. 
In addition, the stripe solution tends to become unstable with the inclusion of the second nearest-neighbor hopping integral ($t'$) 
in the electron dispersion \cite{white99,tohyama99}, although Ref.~\citen{himeda02} reported the opposite. 
The effect of the short- and long-range Coulomb interactions was also explored on a possible 
stabilization of the charge stripes, but no such a tendency was obtained in the comprehensive 
analysis in a large-$N$ theory of the $t$-$J$ model \cite{bejas12,greco16}. 
The situation in the strong coupling Hubbard model is also controversial. State-of-the-art numerical studies 
in various schemes \cite{ccchang10,bxzheng17} showed consistently charge stripes,  but with 
a modulation vector far smaller than the experimental indication \cite{tranquada95}.  
Charge stripes more consistent with the experiment were discussed by combining determinant quantum Monte Calro 
and density matrix renormalization group in the presence of $t'$ \cite{huang18}. 
More extensive studies are required about the stability of the charge stripes as well as the consistency 
between the $t$-$J$ model and the strong coupling Hubbard model, and also about how the stripe order yields 
the nematic state in those models.

The other is to invoke a $d$-wave Pomeranchuk instability ($d$PI) in the metallic phase as was first obtained 
in the $t$-$J$ (Refs.~\citen{yamase00a,yamase00b}) and Hubbard \cite{metzner00} models. 
The $d$PI is equivalent to the $d$-wave bond-charge order at $\vq=(0,0)$ (Ref.~\citen{bejas12}). 
In the present article, we provide theoretical insights into the electronic nematic order 
from a view of the $d$PI.

\subsection{Microscopic origin}
The origin of the nematic order lies in the (effective) nearest-neighbor interaction such as $J$- and $V-$terms 
on a square lattice. We can easily see that the spin-spin interaction contains the following interaction: 
\be
J\sum_{\bra i, j \ket} {\bf S}_i \cdot  {\bf S}_j  \sim 
-\frac{3J}{8N_s}\sum_{\bra i, j \ket} \left( \sum_{\sigma} \tilde{c}_{i \sigma}^{\dagger}\tilde{c}_{j \sigma} \right)
\left( \sum_{\sigma'} \tilde{c}_{j \sigma'}^{\dagger}\tilde{c}_{i \sigma'} \right) \,,
\label{J-dPI}
\ee
where $N_s$ is the total number of lattice sites, 
${\bf S}_i  =\frac{1}{2} \tilde{c}_{i \alpha}^{\dagger} \boldsymbol{\sigma}_{\alpha \beta} \tilde{c}_{i \beta}$ 
with Pauli matrices $\bm{\sigma}$, 
and $\tilde{c}_{i \sigma}^{\dagger}$ ($\tilde{c}_{i \sigma}$) is the creation (annihilation) operator of electrons 
with spin $\sigma$ at site $i$ in the restricted Hilbert space where the double occupancy of electrons  
is prohibited at any site. After the Fourier transformation, we obtain 
\be
{\rm Eq.~(\ref{J-dPI})} = - \frac{3J}{8N_s} \sum_{\vk,\vk',\vq} \sum_{\sigma, \sigma'} 
g(\vk,\vk') \tilde{c}_{\vk-\frac{\vq}{2} \sigma}^{\dagger} \tilde{c}_{\vk+\frac{\vq}{2} \sigma} 
\tilde{c}_{\vk'+\frac{\vq}{2} \sigma'}^{\dagger} \tilde{c}_{\vk'-\frac{\vq}{2} \sigma'}\,. 
\label{J-dPI2}
\ee
Here 
\bea
&&g(\vk,\vk') = \cos (k_x- k_x') +  \cos (k_y- k_y') \\
&&\hspace{12mm} = \frac{1}{2}(\cos k_x + \cos k_y) (\cos k_x' + \cos k_y') \nonumber \\
&& \hspace{15mm} +\frac{1}{2}(\cos k_x - \cos k_y) (\cos k_x' - \cos k_y') \nonumber  \\
&& \hspace{15mm} + \sin k_x \sin k_x' + \sin k_y \sin k_y'  \label{gkk} \,,
\eea
implying four different channels. 

Let us focus on the forward scattering processes and put $\vq={\bf 0}$ in \eq{J-dPI2}; 
see Sec.~3 for a general $\vq \ne {\bf 0}$. 
In this case, the first term in \eq{gkk} describes the $s$-wave channel and corresponds to the so-called 
uniform resonating-valence-bond (RVB) order in the $t$-$J$ model \cite{fczhang88a,lee06}. 
The second term is the $d$-wave channel and describes the nematic order. 
The importance of this $d$-wave channel was not recognized until 2000 \cite{yamase00a,yamase00b}. 
The third and fourth terms are not relevant in the presence of inversion symmetry. 
Therefore the nematic interaction can be extracted from the spin-spin interaction as 
\be
J\sum_{\bra i, j \ket} {\bf S}_i \cdot  {\bf S}_j  \sim 
-\frac{3J}{16 N_s} \sum_{\vk, \vk'} \sum_{\sigma, \sigma'} 
d_{\vk}d_{\vk'} \tilde{c}_{\vk \sigma}^{\dagger} \tilde{c}_{\vk \sigma} 
\tilde{c}_{\vk' \sigma'}^{\dagger} \tilde{c}_{\vk' \sigma'}\,,
\label{dPI}
\ee
where $d_{\vk}=\cos k_x - \cos k_y$. 
The nematic order parameter may be defined as 
\be
\chi_{d} =\frac{1}{N_s} \sum_{\vk, \sigma} d_{\vk} 
\left\bra \tilde{c}_{\vk \sigma}^{\dagger} \tilde{c}_{\vk \sigma} \right\ket \,.
\label{dPIorder}
\ee
When the antiferromagnetic interaction $(J>0)$ is considered, the nematic channel becomes attractive. 
The functional form of the right-hand side of \eq{dPI} is the same as the so-called Landau interaction in the $d$-wave spin-symmetric channel 
if we identify  $\tilde{c}_{\vk \sigma}$ with the usual electron operator $c_{\vk \sigma}$. 
In this sense, the electronic nematic instability described by \eq{dPI} is often called as 
a $d$-wave Pomeranchuk instability ($d$PI), referring to his paper \cite{pomeranchuk59} 
about the stability (not instability) condition  of Fermi liquids. 
Note that the nematic order can occur even without breaking his stability condition because 
the transition can be of first order at low temperature \cite{khavkine04,yamase05}.

Equation~(\ref{dPI}) is not a special feature of the spin-spin interaction. 
It is easy to see that the nearest-neighbor Coulomb interaction ($V>0$) on a square lattice 
contains the same interaction \cite{valenzuela01} 
\be
V\sum_{\bra i, j \ket} \tilde{n}_i \tilde{n}_j \sim - \frac{V}{4N_s} 
\sum_{\vk, \vk'} \sum_{\sigma, \sigma'} 
d_{\vk}d_{\vk'} \tilde{c}_{\vk \sigma}^{\dagger} \tilde{c}_{\vk \sigma} 
\tilde{c}_{\vk' \sigma'}^{\dagger} \tilde{c}_{\vk' \sigma'}\, .
\label{dPI-V}
\ee
If one starts with the Hubbard model, the interaction on the right-hand side of Eqs.~(\ref{dPI}) and (\ref{dPI-V}) 
is generated by electron-electron interactions as a low-energy effective one \cite{metzner00}. 
One can also obtain the $d$PI in a continuum model with central forces \cite{quintanilla06,quintanilla08}. 

Given that the functional form of the Landau interaction is quite general and 
may describe the forward scattering interaction in different electron systems, 
the presence of the $d$-wave Pomeranchuk interaction itself 
is quite general, independent of whether the system is defined in the strong coupling limit as in the $t$-$J$ model 
or in a weak coupling model.

\subsection{Typical phase diagram}
Obviously the spin-spin and Coulomb interactions contain other ordering tendencies, too and 
the nematic order is regarded as one of them. But first let us focus on the nematic order and clarify its typical property. 

\begin{figure}[th]
\centering
\includegraphics[width=7cm]{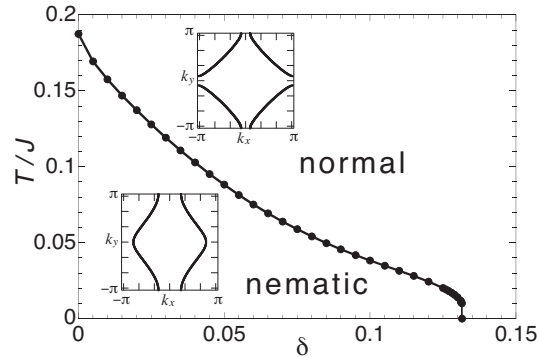}
\caption{Doping dependence of the onset temperature of the nematic instability obtained in the 
slave-boson mean-field theory of the $t$-$J$ model by discarding orders competing with 
the nematic instability. Doping rate $\delta$ is measured from half-filling. 
The typical Fermi surfaces in the normal and nematic phases are shown in each phase. 
Adapted from Ref.~\citen{yamase00b}, where $t/J=4$ and $t'/t=-1/6$ were employed; $t$ and $t'$ are the first and second 
nearest-neighbor hopping on a square lattice, respectively. 
}
\label{dPIphasetJ}
\end{figure}

Figure~\ref{dPIphasetJ} is the phase diagram obtained in the slave-boson mean-field theory 
in the $t$-$J$ model by discarding orders competing with the nematic order. 
 The $d$PI occurs in the low-doping region and is pronounced upon approaching half-filling. 
The energy gain is described by \cite{yamase00b} 
\be
\Delta F \sim \frac{3J}{4} (1-a) (\chi_d)^2 \,,
\ee
and 
\be
a=\frac{3J}{4N_s} \sum_{\vk} d_{\vk}^2 \left( -\frac{\partial f}{\partial \xi_{\vk}} \right)\,.
\ee
Here $f(x)$ is the Fermi distribution function and $\xi_{\vk}$ is the renormalized electron dispersion. 
That is, the nematic order occurs when the $a$ term exceeds unity. 
This $a$ term describes the $d$-wave weighted density of states averaged over 
an energy interval of order of temperature $T$ around the chemical potential 
and becomes large in two different cases. 
One case is that the band width becomes narrower. 
In the slave-boson mean-field theory, 
the nearest-neighbor hopping integral $t$ is renormalized to be $t\delta$, where 
$\delta$ is doping rate measured from half-filling. 
The renormalization of $t$ to $t\delta$ is a special feature of the strong electron correlations that 
the double occupancy of electrons is prohibited at any lattice site. 
This is the major reason  why the onset temperature increases with decreasing doping in \fig{dPIphasetJ}. 
The other case is that the system is close to van Hove filling. 
In \fig{dPIphasetJ} the van Hove filling is located around $\delta=0.10$. The enhancement of the nematic instability 
there is not visible, implying that the effect of the band narrowing is dominant. 
Similar results to \fig{dPIphasetJ} were also obtained 
in the strong coupling Hubbard model \cite{okamoto10,su11,okamoto12}.

For different choices  of band parameters, the phase diagram does not change qualitatively 
close to half-filling. \cite{yamase00b}. 
An additional feature is that van Hove filling can be located in a large doping region and 
the nematic order occurs also around the van Hove filling, 
but with temperature much smaller than \fig{dPIphasetJ} (Refs.~\citen{yamase00b,bejas12}).  
The phase digram becomes similar to that obtained in a weak coupling model except for 
the temperature scale (see \fig{dPIphase} below).

\begin{figure}[th]
\centering
\includegraphics[width=7cm]{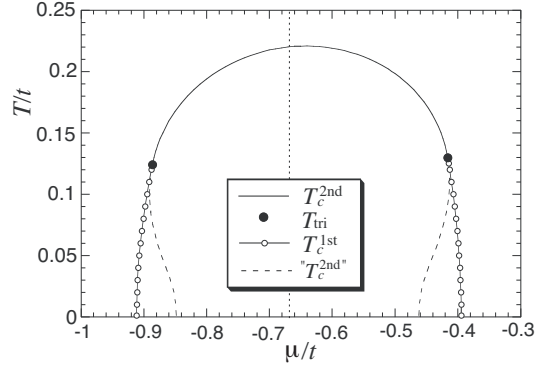}
\caption{Phase diagram of the nematic instability in a weak coupling model, where electrons interact with each other 
via the nematic interaction [see Eqs.~(\ref{dPI}) and (\ref{dPI-V})] with the kinetic energy 
$\xi_{\vk}^{0} = -2 t (\cos k_x + \cos k_y) - 4t' \cos k_x \cos ky -\mu$ with $t'/t=-1/6$ and the chemical potential $\mu$. 
$T_c^{\rm 2nd}$ is a second-order transition line at high temperature and 
$T_c^{\rm 1st}$ describes first-order transition lines at low temperature. 
The end points of the second-order transition line are tricritical points. 
``$T_c^{\rm 2nd}$'' is fictitious second-order transition lines preempted by the first-order transition. 
$\mu=-2/3t$ (dotted line) corresponds to van Hove filling. 
Adapted from Ref.~\citen{yamase05}. 
}
\label{dPIphase}
\end{figure}

In a weak coupling model, the nematic order occurs around van Hove filling with 
a dome-shaped transition line as shown in \fig{dPIphase}: 
a second-order transition on the roof and a first-order transition 
near the edges of the dome \cite{khavkine04,yamase05}. 
The end points of the second-order transition are tricritical points. 

In the weak coupling limit, the phase diagram \fig{dPIphase} is fully determined by a single energy scale  \cite{yamase05}. 
For example, the transition temperature at van Hove filling is given by 
\be
T_{\rm vH}= \frac{2 {\rm e}^{\gamma}}{\pi} \epsilon_{\Lambda} {\rm e}^{-1/(2 \bar{g})} \,,
\label{TcvanHove}
\ee
where $\gamma=0.577$ is the Euler constant, $\bar{g}$ is the dimensionless coupling constant, 
and $\epsilon_{\Lambda}$ corresponds to the typical energy scale around the saddle point contributing to the 
nematic instability. The tricritical temperature is 
\be
T_{\rm tri}={\rm e}^{-\alpha} \epsilon_{\Lambda} {\rm e}^{-1/(2 \bar{g})}
\ee
with $\alpha=0.4515$. Hence the dimensionless ratios of different quantities become universal: 
\be
T_{\rm tri} /T_{\rm vH} = \frac{\pi {\rm e}^{-\alpha}}{2{\rm e}^{\gamma}}=0.5615\,.
\ee
Note that the functional form of \eq{TcvanHove} is exactly the same as $T_c$ in the BCS theory \cite{bardeen57}  
and the presence of universal ratios are also shared with the BCS theory. 

While a certain approximation and some simplification are usually needed to compute the phase diagram, 
we emphasize that 
the presence of the $d$PI interaction in the attractive channel itself does not depend on an approximation. 
Higher order corrections to the approximation may modify quantitative features of the phase diagram, but 
may not introduce a drastic change as long as the nematic instability survives. 
In fact, exact diagonalization \cite{miyanaga06} and variational Monte Carlo \cite{edegger06} studies of the $t$-$J$ model 
suggest qualitatively the same phase diagram as \fig{dPIphasetJ}. 
In a weak coupling model, the effect of nematic fluctuations on the phase diagram 
can be studied in a functional renormalization group scheme. 
Strong fluctuations turn the first order transition into a continuous one, leading to 
a nematic quantum critical point \cite{jakubczyk09}.  
Further strong fluctuations can completely destroy 
the nematic order even at zero temperature in spite of the presence of the attractive interaction \cite{yamase11a}.  
Note that the nematic instability is associated with the breaking of Ising symmetry and thus 
it can occur even at finite temperatures in the two dimensions.

\subsection{Competition with other ordering tendencies}
The electron-electron interaction can also drive other orders such as 
superconductivity, antiferromagnetism, charge-density-wave, and bond-charge orders. 
Which order can be the leading one in the $t$-$J$ and Hubbard models? 
In the $t$-$J$ model, $d$-wave pairing is a stronger instability than the nematic order 
in the slave-boson mean-field theory \cite{yamase00b} and the variational Mote Carlo study \cite{edegger06}.  
Antiferromagnetism also preempts the nematic instability in the slave-boson mean-field theory. 
Exact diagonalization \cite{miyanaga06} found 
a strong tendency of the nematic instability, 
but could not conclude that the nematic order is indeed the leading instability. 
The nematic instability is usually weaker than bond-charge orders in a large-$N$ theory \cite{bejas12,bejas14}, but 
can become a leading one in a heavily doped region when a large $| t' |$ is introduced \cite{bejas12}. 
In the strong coupling Hubbard model, a strong nematic tendency was found 
close to a Mott transition in the cellular dynamical mean-field theory \cite{okamoto10} 
and the dynamical cluster approximations \cite{okamoto10,su11}. 
Spontaneous symmetry breaking to the nematic phase was obtained 
by considering a coupling to the lattice in the cellular dynamical mean-field theory \cite{okamoto12}. 
In the weak coupling Hubbard model, extensive calculations in different functional renormalization group schemes 
showed that the $d$-wave superconductivity and antiferromagnetism are the leading ones \cite{grote02,honerkamp02,kampf03,husemann12} and that charge-density-wave becomes 
a leading order when the sizable nearest-neighbor Coulomb repulsion is taken into account \cite{husemann12}. 
On the other hand, the coexistence of the nematic order and $d$-wave superconductivity 
was found in the second-order perturbation theory \cite{neumayr03} and the dynamical mean-field theory 
combined with the fluctuation exchange method \cite{kitatani17} in the Hubbard model. 
Competition of the nematic instability and superconductivity was studied \cite{yamase07a} in a model, 
which contains only the BCS pairing interaction and the forward scattering interaction such as Eqs.~(\ref{dPI}) and (\ref{dPI-V}).

\subsection{Big response to a small anisotropy} 
\begin{figure}[th]
\centering
\includegraphics[width=7cm]{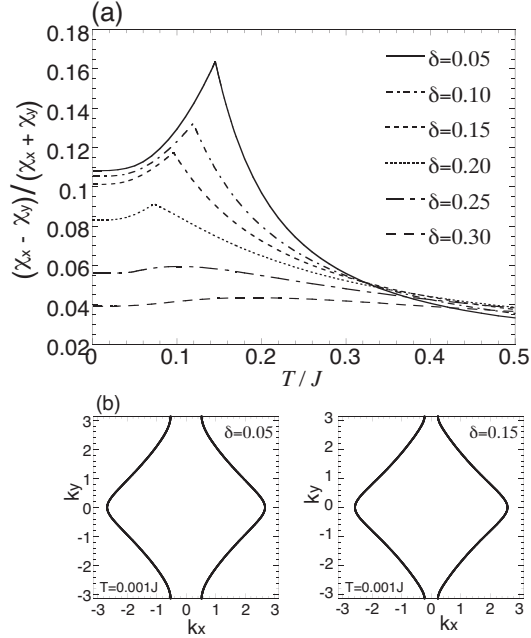}
\caption{(a) Temperature dependence of a degree of anisotropy $\frac{\chi_x-\chi_y}{\chi_x+\chi_y}$ 
for several choices of doping $\delta$ obtained in the slave-boson mean-field theory of 
the $t$-$J$ model with $3\%$ anisotropy. $\chi_{x(y)}$ is the so-called uniform RVB order parameter 
and is defined as $\chi_{\tau} = \langle \sum_{\sigma} f_{i \sigma}^{\dagger}  f_{i+\tau \sigma} \rangle$ with $\tau=x$ 
and $y$;  $\langle \cdots \rangle$ denotes an expectation value; 
no $i$ dependence of $\chi_{\tau}$ is assumed so that it is spatially uniform; $f_{i \sigma}^{\dagger} (f_{i \sigma})$ 
is the creation (annihilation)  operator of slave particles called as spinons with spin $\sigma$ at site $i$.  
(b) Fermi surface deformations due to the nematicity for $\delta=0.05$ and $0.15$ at low temperature.  
From Ref.~\citen{yamase00b}, where $t/J=4$ and $t'/t=-1/6$ were employed. 
}
\label{dPIani}
\end{figure}

Available theoretical results imply that the nematic tendency is surely present in both $t$-$J$ and Hubbard models. 
The point here is that even if the nematic tendency is preempted by a different order, 
the susceptibility of the nematic order can remain large. 
In this case, the electronic property becomes very susceptible to a small external anisotropy and 
exhibits a big anisotropy. 

The big response to a small anisotropy is a robust and a general property of the nematic physics 
when the system is close to the nematic instability. 
This physics was extensively studied in the slave-boson mean-field theory of the $t$-$J$ model 
\cite{yamase00a,yamase00b,yamase01,yamase06,yamase07,yamase09} and 
in the Hubbard model in different schemes:  
the cellular  dynamical mean-field theory \cite{okamoto10,okamoto12} and 
the dynamical cluster approximation \cite{okamoto10,su11}. 

Among various results, we here present  \fig{dPIani}(a), which shows the degree of the anisotropy 
as a function of temperature $T$ for various choices of doping $\delta$ in the $t$-$J$ model with 3 \% anisotropy 
in the nearest-neighbor hopping integral between the $x$ and $y$ directions \cite{yamase00b}. 
At $T=0.5J$  there is a small anisotropy coming from the original external anisotropy. 
With decreasing temperature, the anisotropy is strongly enhanced because of the underlying nematic correlations 
and takes a cusp at the onset temperature of the $d$-wave pairing gap. 
The competition with the pairing formation then suppresses the anisotropy. 
Nevertheless, the big anisotropy remains at zero temperature, which is a few times 
larger than that at high temperature. 
This enhancement is pronounced for lower doping, which is easily expected from \fig{dPIphasetJ}. 
Consequently the shape of the Fermi surface can be deformed substantially as shown in \fig{dPIani}(b). 
Since the external anisotropy is rather small in \fig{dPIani}(a), the curves in \fig{dPIani}(a) 
are regarded as the temperature dependence of the nematic susceptibility. 

In the theoretical scheme in Ref.~\citen{yamase00b}, the nematic tendency increases with decreasing doping 
as shown in Figs.~\ref{dPIphasetJ} and \ref{dPIani}(a). 
However, results in the vicinity of $\delta=0$ should be taken carefully because the Mott physics is considered 
mainly as the band narrowing effect in the slave-boson mean-field theory of the $t$-$J$ model. 
Emergence of antiferromagnetism is not considered near half-filling. 
As described in Sec.~2.3, the nematic order competes with 
other ordering tendencies and thus would be suppressed eventually in the vicinity of half-filling. 
The strong coupling Hubbard model with a large on-site Coulomb interaction $U$ 
($U/t=6, 8, 10, 12$ in Ref.~\citen{okamoto10}, 
$U/t=6$ in Ref.~\citen{su11}, and $U/t=10$ in Ref.~\citen{okamoto12}) also shows the large nematicity 
in a small doping region close to half-filling, consistent with Figs.~\ref{dPIphasetJ} and \ref{dPIani}(a). 
However, it should be kept in mind that the possible magnetic instability was not considered in those studies.

\subsection{Spectral function} 
Hereafter we do not distinguish between electron operators $\tilde{c}_{\vk \sigma}$ and $c_{\vk \sigma}$ 
because the nematic physics is relevant to both strong and weak coupling models. 
The nematic correlation function is defined as 
\be
\kappa_{d}(\vq, \omega) = \frac{i}{N_s} \int_{0}^{\infty} d t \;
\bra [\hat{\chi}_d(\vq, t),  \hat{\chi}_d(-\vq, 0)] \ket  {\rm e}^{i (\omega + i \Gamma) t}  \,,
\label{kappad}
\ee
where $\vq$ and $\omega$ are momentum and energy transfer, respectively,  
$\Gamma$ is a positive infinitesimal,  
$\hat{\chi}_d (\vq) = \frac{1}{N_s} \sum_{\vk \sigma} d_{\vk} c_{\vk-\frac{\vq}{2} \sigma}^{\dagger} 
c_{\vk+\frac{\vq}{2}  \sigma}$ is a generalized nematic operator [see also \eq{dPIorder}], 
$\hat{\chi}_{d}(\vq, t) = {\rm e}^{i \mathcal{H} t} \hat{\chi}_{d}(\vq)  {\rm e}^{-i \mathcal{H} t}$ 
for the Hamiltonian $\mathcal{H}$, $[ \cdot, \cdot ]$ is the commutator, 
and $\bra \cdots \ket$ denotes an expectation value under the 
Hamiltonian $\mathcal{H}$. 

\begin{figure}[th]
\centering
\includegraphics[width=12cm]{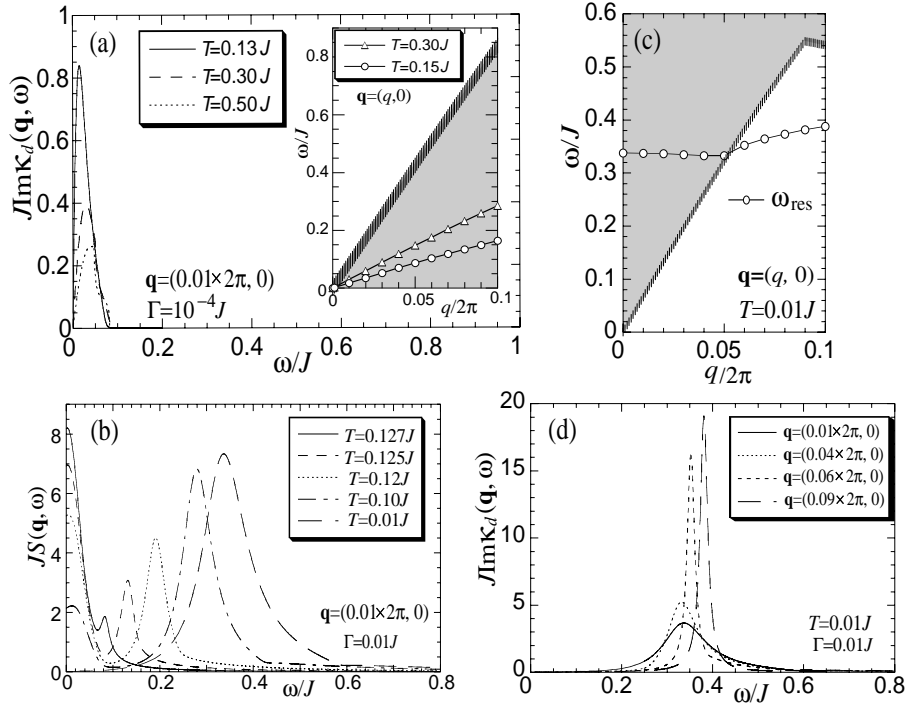}
\caption{(a) Spectral function of nematic fluctuations for momentum $\vq$ close to $(0,0)$ 
for several choices of temperatures in the normal phase (a) and the $d$-wave pairing state (b). 
Inset in (a) shows the excitation spectrum on the plane of energy $\omega$ and $\vq$. 
The shaded region is a particle-hole continuum and the open symbols correspond to the peak position 
of Im$\kappa_{d}(\vq,\omega)$. 
In (b), $S(\vq,\omega) = 2 {\rm Im}\kappa_{d}(\vq,\omega)/(1-{\rm e}^{-\omega/T})$ is plotted 
to highlight the low-energy structure. 
(c) Excitation spectrum in the $d$-wave pairing state. The shaded region is a particle-hole continuum. 
The open circle, $\omega_{\rm res}$, corresponds to the peak position of Im$\kappa_{d}(\vq,\omega)$. 
(d) Spectral function of nematic fluctuations for different values of $\vq$ at low temperature. 
Adapted from Ref.~\citen{yamase04b}, where the slave-boson scheme of the $t$-$J$ model was 
employed for $t/J=4$ and $t'/t=-1/6$. 
}
\label{dPIspec}
\end{figure}

The spectrum of the nematic fluctuations was revealed in Ref.~\citen{yamase04b} in both 
normal and superconducting phases. 
In the normal phase, the nematic mode is realized inside the particle-hole continuum [inset in \fig{dPIspec}(a)].  
Upon approaching the nematic instability with decreasing temperature, 
the velocity of the nematic mode decreases, accompanied by the strong enhancement of the low-energy peak [\fig{dPIspec}(a)]. 
However, pairing instability preempts the nematic instability below $T=0.128J$. 
The low-energy spectral weight is gradually transferred to high energy 
as shown in \fig{dPIspec}(b) and a peak is realized at a finite energy. 
At the same time, the pairing formation generates a gap  
and the continuum is now realized on the upper side as shown in the gray region in \fig{dPIspec}(c). 
When $\vq$ is small, the peak is rather broad [\fig{dPIspec}(d)] because of the mixture 
of the particle-hole continuum. 
However, at an intermediate value of $\vq$, a very sharp peak is realized [\fig{dPIspec}(d)]. 
This is the resonance mode of the nematic fluctuations and is located inside a gap as shown in \fig{dPIspec}(c). 
While the nematic mode has a linear dispersion in \fig{dPIspec}(a), it becomes a gapped mode with a rather flat dispersion 
as a function of $\vq$ inside the superconducting phase [\fig{dPIspec}(c)].   
The predicted resonance mode has not yet been reported in cuprates 
despite many experimental indications of nematic correlations \cite{hinkov04,hinkov07,hinkov08,haug10,daou10,cyr-choiniere15,sato17,nakata18,auvray19,yamase21}.  

The linear dispersion shown in the inset in \fig{dPIspec}(a) should not be confused with the zero-sound mode, which 
is also characterized by a gapless linear dispersion near $\vq=(0,0)$. 
The zero-sound mode is driven by a short-range {\it repulsive} interaction and is realized {\it above} the particle-hole continuum \cite{negele}. On the other hand, the nematic mode originates from a short-range {\it attractive} interaction in a $d$-wave channel [see Eqs.~(\ref{J-dPI2}) and ~(\ref{gkk})] and is realized {\it inside} the particle-hole continuum in the normal phase.

\subsection{Electron self-energy from nematic fluctuations} 
In the normal phase, the nematic fluctuations appear in a low-energy region as shown in \fig{dPIspec}(a). 
Its propagator can take the form in general \cite{dellanna06,yamase12}
\be
 D_{\vk\vk'}(\vq,\nu_n) =
 \frac{g \, d_{\vk} d_{\vk'}}
 {(\xi_0/\xi)^2 + \xi_0^2 |\vq|^2 + |\nu_n|/(u|\vq|)} \; ,
\label{dPI-D}
\ee
where $\nu_n = 2\pi n T$ is a bosonic Matsubara frequency; $\xi$ 
is the nematic correlation length, $\xi_0$ and $u$ are 
non-universal parameters determined by the momentum dependence 
of the interaction strength and the band structure; $g$ is a coupling strength at $\vq=(0,0)$.

Quantum nematic fluctuations are described by finite Matsubara frequencies $\nu_{n} \ne 0$ in \eq{dPI-D}. 
References~\citen{dellanna06} and \citen{metzner03} studied how they renormalized the one-particle property of electrons.   
Non-Fermi liquid behavior was obtained at a quantum critical point, otherwise the Fermi liquid state was stabilized 
in the ground state. In both cases, the spectral function for one-particle excitations $A(\vk,\omega)$ exhibits 
a single peak at $\vk=\vk_{\rm F}$ and $\omega=0$; $\vk_{\rm F}$ is the Fermi momentum. 

At finite temperature, a more drastic effect occurs \cite{yamase12}. 
Since the quantum fluctuations are cut off by temperature, 
we may focus on the thermal fluctuations in a relatively high temperature region and put $\nu_n=0$ in \eq{dPI-D}. 

\begin{figure}[th]
\centering
\includegraphics[width=10cm]{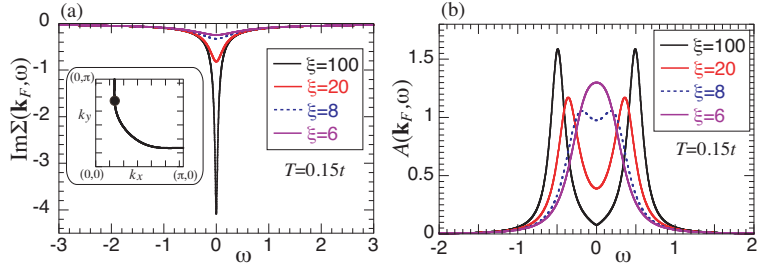}
\caption{(color online) Electron self-energy (a) and spectral function (b) obtained 
in the perturbative first-order calculation of the self-energy from thermal nematic fluctuations  
for several choices of the nematic correlation length $\xi$ at temperature $T=0.15t$; 
the lattice constant is set to unity. 
The momentum is chosen at the Fermi momentum depicted in the inset in (a). 
The original result (b) is given in Ref.~\citen{yamase12}. 
}
\label{1stAkw}
\end{figure}

The self-energy computed perturbatively to first order is shown in \fig{1stAkw}(a) for several choices of 
the nematic correlation length $\xi$. 
The self-energy exhibits a peak at $\omega=0$ for the momentum on the Fermi surface 
and the peak develops to be sharper with increasing $\xi$.  
Consequently, as shown in \fig{1stAkw}(b),  
a peak of the spectral function $A(\vk_{F},\omega)$ at $\omega=0$ splits and forms  
a double peak structure with strong suppression of the spectral weight at $\omega=0$, 
indicating a gaplike feature around the Fermi surface, although the system is in the disordered phase. 
We refer to this gaplike feature as a pseudogap, which is similar to the pseudogap observed in cuprates \cite{timusk99}. 

\begin{figure}[th]
\centering
\includegraphics[width=15cm]{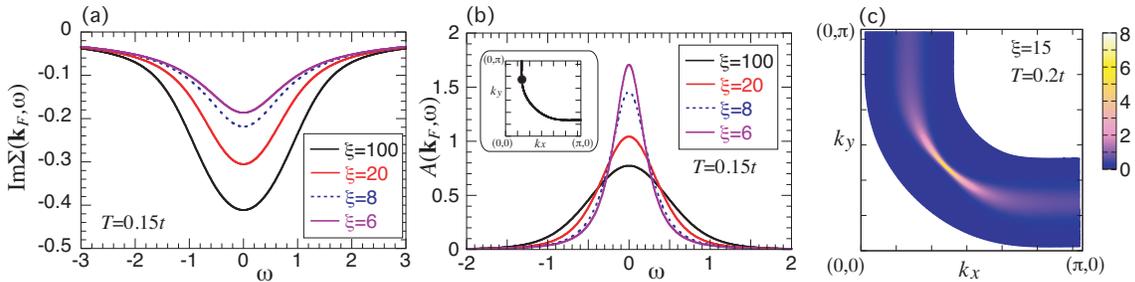}
\caption{(color online) Electron self-energy (a) and spectral function (b) obtained 
in a nonperturbative resummation of contributions from thermal nematic fluctuations 
for several choices of the nematic correlation length $\xi$ at temperature $T=0.15t$; 
the lattice constant is set to unity. The momentum is chosen 
at the Fermi momentum depicted in the inset in (b). 
(c) Intensity map of the spectral function $A(\vk,0)$ in the first quadrant of the Brillouin zone 
for the nematic correlation length $\xi=15$ at $T=0.2t$.  
The original results (b) and (c) are given in Ref.~\citen{yamase12}. 
}
\label{Akw}
\end{figure}

The results in \fig{1stAkw} suggest a pseudogap driven by the thermal nematic fluctuations. 
However, when more precise calculations were performed by including the self-consistency in the 
perturbative analysis and also vertex corrections \cite{yamase12}, 
the pseudogap feature in \fig{1stAkw}(b) disappears. 
The obtained self-energy is shown in Fig.~\ref{Akw}(a). 
It still exhibits a peak at $\omega=0$ for $\vk = \vk_{F}$ and 
the peak is pronounced more upon increasing $\xi$. 
But compared with \fig{1stAkw}(a), the absolute value of Im$\Sigma(\vk_{F},\omega)$ is suppressed substantially. 
Consequently the spectral function forms a single peak at $\omega=0$ as shown in \fig{Akw}(b), where 
there is no indication of the gaplike feature 
with increasing $\xi$, although the peak at $\omega=0$ is suppressed. 

In \fig{Akw}(b) the peak width is proportional to $\sqrt{\log \xi}$ and is broadened with increasing $\xi$. 
This feature is very different from that from the quantum nematic fluctuations. 
In the quantum critical regime \cite{dellanna06}, 
the spectral function also exhibits a single peak, but with the peak width proportional to 
$T \xi$; here $\xi$ diverges as $( T | \log T | )^{-1/2}$ upon approaching the quantum critical point at $T=0$. 
The single peak structure in \fig{Akw}(b) should not be understood as the indication of a Fermi liquid. 
Rather it can be interpreted as an incoherent peak in the sense that 
the self-energy exhibits a peak at $\omega=0$ as shown in \fig{Akw}(a). 

Figure~\ref{Akw}(c) is a map of the spectral function at $\omega=0$ in the first quadrant of the Brillouin zone. 
While the nematic correlation length $\xi$ is rather small, the spectral weight is substantially suppressed 
near $\vk=(\pi,0)$ and exhibits a Fermi-arc-like feature, although no gaplike feature is realized around $\vk=(\pi,0)$ 
as seen in \fig{Akw}(b).

\subsection{Probes of nematicity} 
\subsubsection{Raman scattering and ultrasound} 
\begin{figure}[th]
\centering
\includegraphics[width=12cm]{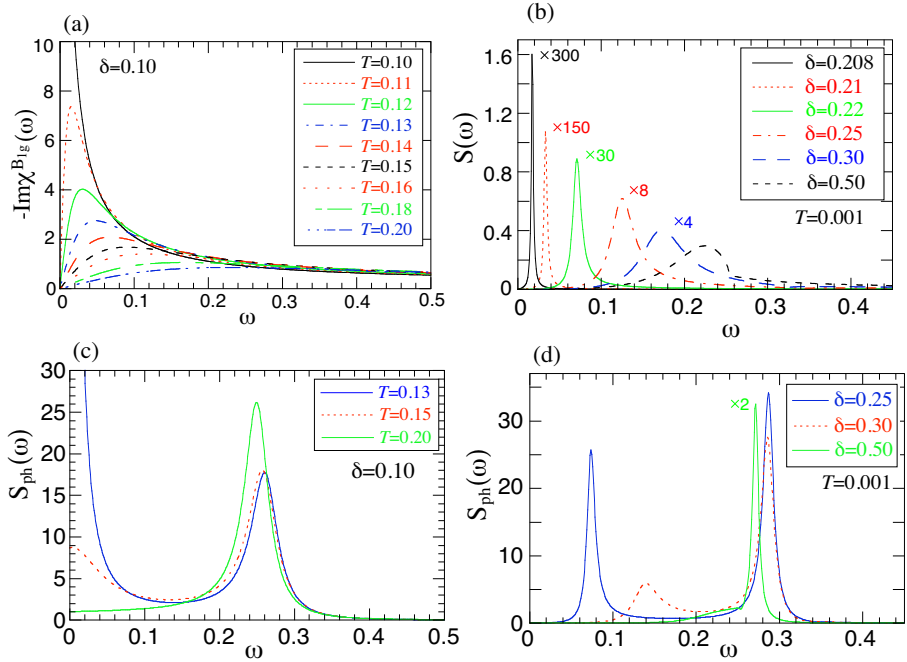}
\caption{(color online) 
(a) Electronic Raman intensity in the $B_{1g}$ channel for a sequence of temperatures $T$ 
close to the nematic instability $T_c=0.0984$ in the normal phase at $\delta=0.10$; 
the energy unit is $t$. 
(b) Electronic Raman intensity in the $B_{1g}$ channel for a sequence of doping concentrations $\delta$ 
upon approaching the nematic quantum critical point $\delta=0.207$ 
inside the $d$-wave superconducting phase. The results are presented for $S(\omega) = 
- {\rm Im}\chi^{\rm B_{1g}}(\omega) /(1-{\rm e}^{-\omega/T})/\pi$. 
The actual value of $S(\omega)$ is obtained by multiplication with the factor indicated near each peak 
except for $\delta=0.50$. 
(c) Raman intensity from $B_{1g}$ phonon scattering for several choices of temperatures  
close to the nematic instability at $T_c=0.126$ in the normal phase at $\delta=0.10$. 
(d) Raman intensity from $B_{1g}$ phonon scattering for several choices of doping concentrations 
upon approaching the nematic quantum critical point $\delta_c=0.233$ inside the $d$-wave 
superconducting phase. The actual intensity at $\delta=0.50$ is obtained by multiplying by 
a factor of $2$.  Adapted from Ref.~\citen{yamase11}, where electrons interact with each other 
via the forward scattering channel in the nematic interaction [Eqs.~(\ref{dPI}) and (\ref{dPI-V})] and 
also with phonons; the electron self-energy was also considered in the normal phase by modeling 
the bosonic spectral function $\alpha^{2} F$ phenomenologically. 
}
\label{Raman}
\end{figure}

Electronic Raman scattering in the $B_{1g}$ symmetry measures directly 
the nematic correlation function for $\vq=0$ in \eq{kappad}  
and can provide microscopic evidence of nematic fluctuations in real materials \cite{yamase11}.  
When the system approaches the nematic instability with decreasing temperature, 
the so-called central peak was predicted [\fig{Raman}(a)]. 
On the other hand, when the nematic quantum critical point is hidden inside the superconducting phase  
and the system approaches it upon changing the doping rate, the softening of the $B_{1g}$ peak was  
predicted [\fig{Raman}(b)]. 

$B_{1g}$ phonons also couple directly to the nematic fluctuations and Raman scattering exhibits 
a characteristic feature near the nematic instability \cite{yamase11}. A caveat here is the $B_{1g}$  phonon energy 
is assumed to be relatively large, e.g., around 40 meV in YBa$_{2}$Cu$_{3}$O$_{6+y}$ (YBCO). 
In this case, the original phonon frequency 
itself does not change much and stays around $\omega \approx 0.25$ in \fig{Raman}(c) and $0.28$ in \fig{Raman}(d). 
Instead, a central peak emerges in the phonon spectrum in the normal phase 
[\fig{Raman}(c)] and a soft phonon mode in the superconducting phase [\fig{Raman}(d)]. 

When the $B_{1g}$ phonon's frequency is very small, the predicted double peak structure 
in the normal phase in \fig{Raman}(c) may overlap with each other and look like a single peak. 
In the superconducting state, the original phonon mode simply softens down to zero energy and no 
additional low-energy peak emerges.

In Bi$_2$Sr$_2$CaCu$_2$O$_{8+\delta}$ (Bi2212) around van Hove filling, 
where the pseudogap temperature nearly vanishes, 
the $B_{1g}$ Raman scattering \cite{auvray19} showed a central peak as predicted in \fig{Raman}(a). 
The nematic correlations are then suppressed inside the pseudogap region. 
Further exploration by Raman scattering for different cuprate compounds 
will serve to elucidate how the nematic fluctuations evolve in the cuprate phase diagram \cite{keimer15}. 

The theory of Raman scattering from nematic fluctuations  
can be applied to other materials \cite{yamase13a}. 
In iron-based superconductors  extensive Raman scattering measurements 
reported the emergence of the central mode in the normal state [\fig{Raman}(a)] 
(Refs.~\citen{gallais13,thorsmolle16,massat16,sfwu17}) 
and an in-gap mode in the superconducting state [\fig{Raman}(b)] (Refs.~\citen{thorsmolle16,sfwu17}). 
Similar data, however, can be interpreted in different scenarios invoking the spin sector: 
manifestation of spin nematic fluctuations \cite{kretzschmar16} and 
frustration of localized spins \cite{baum19}. 

Ultrasound waves also couple to the nematic fluctuations. In particular, it was pointed out theoretically that 
the nematic fluctuations enhance the transverse sound attenuation and sound-velocity softening along the $[110]$ direction 
upon approaching the nematic instability whereas they remain unaffected along the $[100]$ direction \cite{adachi09}.  
The transverse phonons along the $[110]$ direction have the electron-phonon vertex characterized 
by $B_{1g}$ symmetry, the same symmetry as the $B_{1g}$ phonon discussed above. 
We are not aware of experimental tests of the ultrasound wave anomaly in cuprates.

\subsubsection{ARPES and Compton scattering} 
As seen in the original theoretical finding of the $d$PI \cite{yamase00a,yamase00b,metzner00}, 
the nematic instability deforms the Fermi surface (FS). 
Hence the observation of a Fermi surface deformation can be the direct evidence of the nematic order. 
Recent angle-resolved photoemission spectroscopy (ARPES) measurements for Bi2212 
reported the enhancement of the band anisotropy 
when applying a uniaxial strain to the material \cite{nakata18}, similar to the theoretical prediction shown in \fig{dPIani}. 

Compton scattering measures directly the momentum distribution function and thus can also detect the FS, 
complementary to ARPES. Recent measurements for La$_{2-x}$Sr$_x$CuO$_4$ (LSCO) 
at $x=0.08$, $0.15$, and $0.30$ revealed that 
the FS strongly deforms to become open for $x=0.08$ and $0.15$, like a quasi-one-dimensional FS, 
in each CuO$_2$ plane and such a deformation alternates along the $c$-axis, recovering the fourfold symmetry 
in bulk \cite{yamase21}. 
The nematicity is most pronounced at $x=0.08$, decreases with increasing doping, and nearly vanishes at $x=0.30$, 
consistent with the theoretical prediction (Figs.~\ref{dPIphasetJ} and \ref{dPIani}).

\subsubsection{xy-anisotropy of physical quantities} 
YBCO has an intrinsic small anisotropy coming from the orthorhombic  crystal structure. 
Various physical quantities, however, can exhibit a large $xy$-anisotropy due to a coupling to 
the underlying nematicity, which is a robust feature of the nematic physics as was discussed in Fig.~\ref{dPIani}. 

The first experimental signature was obtained in the $xy$-anisotropy of the magnetic excitation spectra 
in YBCO \cite{hinkov04,hinkov07,hinkov08,haug10}. 
While the anisotropy is observed in the magnetic excitations, this does not necessarily mean that 
the origin of the nematicity should lie in the magnetic sectors such as spin nematic \cite{blume69,andreev84,orth19}. 
In fact, the observed anisotropy was well captured in terms of the underlying nematic correlations from the $d$PI  
in the $t$-$J$ model \cite{yamase06,yamase09}. 

The Nernst coefficient also exhibits the strong $xy$-anisotropy,  
which seems pronounced below the pseudogap temperature \cite{daou10} or 
below a certain temperature inside the pseudogap phase \cite{cyr-choiniere15}. 
The large anisotropy of the Nernst coefficient can be undertood as a consequence of 
Fermi surface distortions due to nematicity \cite{hackl09}.  

Magnetic torque measurements for YBCO reported that a component of the two-fold symmetry 
starts to enhance across the pseudogap temperature and the authors argued that 
the pseudogap temperature corresponds to the nematic instability \cite{sato17}. 
If this is indeed so, the electronic Raman scattering in $B_{1g}$ symmetry 
should exhibit a central peak close to the pseudogap temperature [see \fig{Raman}(a)].  
On the other hand, in contrast to \fig{1stAkw}(b), the elaborate calculations in \fig{Akw} found that 
the nematic fluctuations do not generate a gap along 
the FS, although the Fermi-arc-like feature is generated.  
In addition, the experimental data \cite{sato17} can also be interpreted 
in terms of antiferromagnetic fluctuations with the Dzyaloshinskii-Moriya interaction 
without invoking the nematic physics \cite{morinari18}. 
The magnetic torque results will stimulate further studies on 
the relationship between the nematic order and the pseudogap.

\subsection{Global view of the nematic physics: a concept of Griffiths wings.}  
\begin{figure}[th]
\centering
\includegraphics[width=8cm]{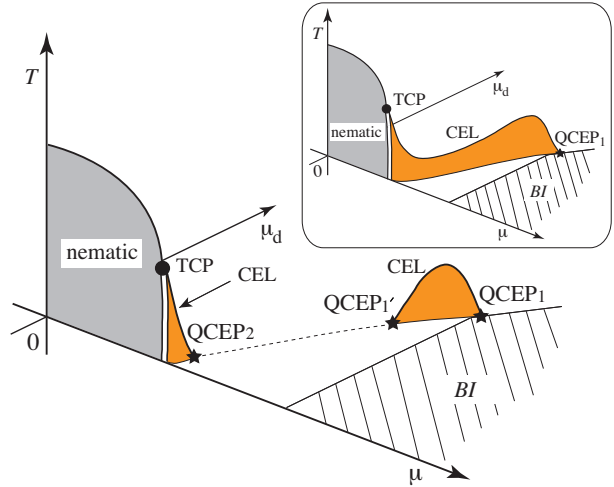}
\caption{(color online) Schematic phase diagram of the nematic phase transition in the presence of 
weak nematic order-parameter fluctuations in the plane of anisotropy 
$\mu_d$, the chemical potential $\mu$, and temperature $T$; 
the nematic interaction in the $d$-wave channel [see Eqs.~(\ref{J-dPI2}) and (\ref{gkk})] was considered 
as the electron-electron interaction. No anisotropy is present at $\mu_d=0$. 
$\mu=0$ may correspond to half-filling 
or van Hove filling. The transition is of second order at high temperature (solid line) and 
of first order at low temperature (double line). The solid circle denotes the TCP. 
The phase diagram in the $\mu$-$T$ plane at $\mu_d=0$ is similar to Figs.~\ref{dPIphasetJ} and \ref{dPIphase}. 
The band insulating (BI) state is realized in the  striped region.  The wings (colored in orange) describe  
first-order transition surfaces, where the nematicity changes discontinuously 
(the meta-nematic transition). The stars denote quantum critical end points (QCEP). 
The inset is the corresponding phase diagram obtained in mean-field theory, 
where no nematic fluctuations are present. Adapted from Ref.~\citen{yamase15}. 
}
\label{griffiths}
\end{figure}

The nematic order couples directly to an external $xy$ anisotropy such as strain, uniaxial pressure, 
and an orthorhombic crystal structure. In the presence of $xy$ anisotropy, the nematic order parameter 
becomes finite and thus no second-order phase transition occurs. However, a first-order phase transition 
is still possible. The nematic phase diagram was revealed in the three-dimensional space spanned 
by the chemical potential $\mu$, the external $xy$ anisotropy $\mu_d$, and temperature $T$ (Ref.~\citen{yamase15}).  
The inset in \fig{griffiths} is a mean-field phase diagram. 
The phase diagram is symmetric with respect to the axis $\mu_d=0$ and 
almost symmetric with respect to the axis of $\mu=0$ as long as $| t' |$ is small. 
A wing (colored in orange) develops from the first-order phase transition line at $\mu_d=0$. 
Crossing this wing, the nematicity changes discontinuously, that is, a meta-nematic transition occurs. 
This is a first-order nematic phase transition. 
The top edge of the wing corresponds to  the critical end line (CEL) and becomes a tricritical point (TCP) 
at $\mu_d=0$. The wing disappears at a quantum critical end point (QCEP) near the edge of the band insulator.

When weak nematic order-parameter fluctuations are included, the wing splits into two 
as shown in the main panel in \fig{griffiths}.  
One wing is realized in a large anisotropy region and the other is in a region close to zero anisotropy. 
In this case, a QCEP is realized close to the tetragonal phase ($\mu_d=0$) and thus 
strong nematic fluctuations are expected even if the external anisotropy is present. 
No phase transition is present between the two wings depicted by the dashed line in \fig{griffiths}. 

At zero temperature, YBCO is expected to locate near the point of QCEP$_2$ in \fig{griffiths} and 
$\mu=0$ may correspond to half-filling. In fact, the magnetic excitation spectra exhibit the pronounced anisotropy 
in  YBCO$_{6.3}$,  YBCO$_{6.35}$, and YBCO$_{6.45}$ (Refs.~\citen{hinkov08,haug10}) 
whereas the anisotropy becomes moderate in YBCO$_{6.6}$ (Ref.~\citen{hinkov07}) 
and YBCO$_{6.85}$ (Ref.~\citen{hinkov04}). This suggests that the QCEP$_2$ in \fig{griffiths} 
corresponds to a carrier density between YBCO$_{6.45}$ and YBCO$_{6.6}$. 

Referring to a pioneering work by Griffiths about the wing structure associated with a first-order phase transition, 
namely a concept of tricritical point, in the He$^3$-He$^4$ mixtures \cite{griffiths70}, 
the wing structure in \fig{griffiths} may be 
called as Griffiths wings associated with the nematic transition. 
Griffiths wings are also known in metallic ferromagnetic systems and were confirmed in UGe$_2$ (Ref.~\citen{kotegawa11})  
and UCoAl (Ref.~\citen{aoki11}).

\section{Bond-charge orders and their fluctuations} 
The possible importance of bond-charge orders in cuprates was already recognized in early theoretical studies \cite{affleck88,marston89,sachdev91,vojta99,vojta02}. 
The RVB theory of the $t$-$J$ model was formulated also 
by introducing bond-order parameters \cite{fczhang88a,lee06}. 
Recent resonant x-ray scattering (RXS) \cite{da-silva-neto15,da-silva-neto16} 
and resonant inelastic x-ray scattering (RIXS) \cite{da-silva-neto18} measurements suggest a 
bond-charge order, which however seems different from what was discussed in early studies. 

Extensive studies of possible charge orders were performed in a large-$N$ theory of 
the $t$-$J$ model on a square lattice at leading order \cite{bejas12,bejas14}. 
This theory has an advantage to study all possible charge instabilities present in the $t$-$J$ model 
on an equal footing. 
The theory can be formulated in different schemes \cite{morse91,cappelluti99}  
including a path integral representation \cite{foussats04}.   
The path integral formalism was shown to yield results consistent with 
exact diagonalization \cite{bejas06}. 
We review theoretical insights obtained in that formalism focusing on bond-charge orders in this section.  
Usual on-site charge fluctuations shall be reviewed in the next section by including the long-range Coulomb interaction.

In the leading-order theory of the large-$N$ expansion, 
the number of spins is extended from 2 to $N$, and physical quantities are computed 
at the order of $1/N$. 
The charge susceptibilities $D_{ab}(\vq, \omega)$ are obtained as a $6 \times 6$ matrix and given by  
\be
[D_{ab}(\vq, \omega)]^{-1} 
= [D^{(0)}_{ab}(\vq, \omega)]^{-1} - \Pi_{ab}(\vq, \omega)\,, 
\label{dyson}
\ee
where $a$ and $b$ run from 1 to 6; $\vq$ and $\omega$ are momentum and energy transfer, respectively.  
The quantity $D^{(0)}_{ab}$ 
describes bare charge susceptibilities and 
is renormalized by the boson self-energies $\Pi_{ab}$ at the order of $1/N$. 
Explicit forms of $D^{(0)}_{ab}$ and $\Pi_{ab}$ are given in Ref.~\citen{bejas12}. 
In this scheme, the tendency toward phase separation becomes rather strong and thus the nearest-neighbor 
Coulomb interaction is usually introduced to avoid the phase separation. 
This does not introduce essential changes in the underlying tendency of various bond orders as well as 
their excitation spectra.

\subsection{Bond-charge orders} 
\begin{figure}[th]
\centering
\includegraphics[width=14cm]{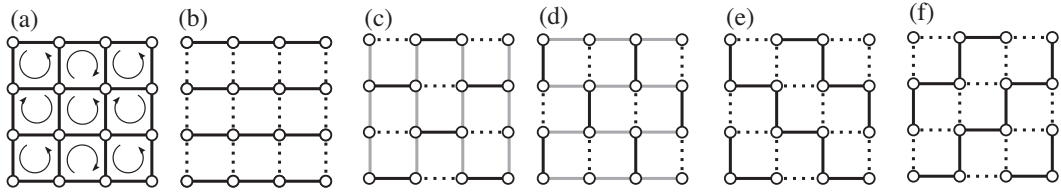}
\caption{Sketch of various bond-charge orders in real space obtained in the large-$N$ theory of the $t$-$J$ model. 
The black and dotted lines denote a stronger and weaker bond, respectively; 
the gray line  corresponds to the intermediate strength of the bond. 
(a) Flux phase with $\vq=(\pi,\pi)$, where staggered circulating currents flow in each plaquette. 
(b) Nematic state driven by $d$PI. 
In momentum space the Fermi surface deforms as shown in \fig{dPIphasetJ}. 
(c) and (d) Uniaxial state with modulation vector $\pi$ 
along the $x$ and $y$ direction, respectively. 
(e) and (f) $s$-wave and $d$-wave state with $\vq=(\pi,\pi)$, respectively. 
The $d$-wave state with $\vq=(0,0)$ is equivalent to the nematic state and 
that with $\vq \sim (0.5\pi, 0)$ is relevant to the charge order tendency observed 
in electron-doped cuprates \cite{da-silva-neto15,da-silva-neto16,da-silva-neto18}; see also \fig{bondspectrum}.  
Adapted from Ref.~\citen{bejas14}. 
}
\label{bondorder}
\end{figure}

The instability of the paramagnetic phase is signaled by the divergence of the static susceptibilities 
$D_{a b}(\vq, 0)$ for a continuous phase transition. Therefore we study eigenvalues and eigenvectors 
of the matrix $[D_{a b} (\vq, 0)]^{-1}$. When an eigenvalue crosses zero at a given doping rate, temperature, and momentum, 
the instability occurs toward a phase characterized by the corresponding eigenvector. 
Among numerous possibilities, essentially only a few bond-charge instabilities are relevant to 
parameters appropriate to cuprates. 
i) Flux phase with $\vq \approx (\pi, \pi)$. In this phase, currents flow in each plaquette; see \fig{bondorder}(a). 
ii) the $d$PI [see \fig{bondorder}(b)]. 
The $d$PI leads to the electronic nematic state as already described in Sec.~2. The $d$PI can have a finite 
momentum $\vq$ near $(0,0)$, which is often referred to as an incommensurate $d$PI \cite{metlitski10,metlitski10a,holder12,husemann12,sachdev13}. 
iii) Various bond-charge orders with $\vq \approx (\pi,\pi)$: 
uniaxial bond-charge order (unibond)  [Figs.~\ref{bondorder}(c) and (d)], 
$s$-wave bond-charge order ($s$bond) [\fig{bondorder}(e)], and  
$d$-wave bond-charge order ($d$bond) [\fig{bondorder}(f)]. 
The unibond has a bond amplitude modulated only along the $x$ or $y$ direction, whereas $s$bond and $d$bond 
with $\vq = (\pi,\pi)$ have a bond amplitude modulated along both $x$ and $y$ directions, and its amplitude is 
inphase and antiphase, respectively. 
The $d$bond with $\vq=(0,0)$ is the same as the $d$PI. 
That is, the $d$PI is a special type of the $d$bond among various types of bond orders. 

\begin{figure}[th]
\centering
\includegraphics[width=14cm]{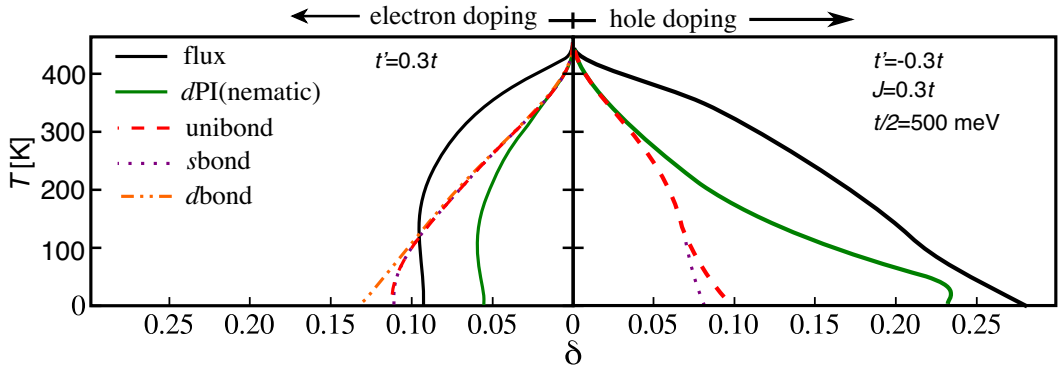}
\caption{(color online) 
Phase digram of charge orders obtained in the large-$N$ theory of the $t$-$J$ models. 
The original results are given in Ref.~\citen{bejas14}, where 
$t=500\; {\rm meV}$, instead of $t/2=500\; {\rm meV}$, was invoked. 
}
\label{bondphase}
\end{figure}

The phase diagram obtained in the large-$N$ theory of the $t$-$J$ model \cite{bejas14} 
is shown in \fig{bondphase} for both hole- and electron-doped cases. 
At half-filling, flux phase, $d$PI, $d$bond, $s$bond, and unibond have the same onset temperature $T_c=J/8$. 
 Upon carrier doping such a degeneracy is lifted. 
 On the hole-doped side, flux phase with $\vq=(\pi,\pi)$ is the leading instability and 
 the $d$PI with $\vq \approx (0,0)$ is the second leading one; when a larger $|t'|$ is taken, 
 the $d$PI would extend to a higher doping region and become the leading one there \cite{bejas12}. 
 The $s$bond and unibond  have ordering tendencies around $\vq=(\pi,\pi)$, which 
 are nearly degenerate and much weaker than flux phase and the $d$PI. 
 
 The charge ordering tendency exhibits a strong particle-hole asymmetry. 
 In contrast to the hole-doped case, 
 the charge ordering tendency is limited only close to half-filling in the electron doping region. 
 The $d$PI becomes the weakest instability and flux phase is leading at high temperature. 
 At low temperature in a moderate doping region, $d$bond with $\vq \approx (0.8\pi, 0.8\pi)$ becomes dominant 
 and the ordering tendency of $s$bond and unibond with $\vq \approx (\pi, \pi)$ are located  close to the $d$bond. 
 
 In the large-$N$ theory, the effect of spin fluctuations does not appear at the leading order and thus 
 no magnetic instability is observed in \fig{bondphase}. 
 While this is an advantage of the large-$N$ theory in that one can focus on the charge degree of freedom  
 in the $t$-$J$ model, a comparison with experiments should be made carefully 
 by keeping in mind a possible magnetic instability. 

 On the hole-doped side, one may assume the magnetic order close to half-filling, for example, in $\delta \lesssim 0.05$. 
 Hence the theory predicts the flux instability in a large doping region, which is however not observed in experiemnts. 
 This inconsistency remains to be studied. 
 Since the second leading one is the $d$PI, we can expect that the systems has a large nematic susceptibility and becomes 
 sensitive to an external $xy$ anisotropy even if the $d$PI does not occurs. 
 This feature is in fact consistent with the nematic tendency observed 
 in cuprates \cite{hinkov04,hinkov07,hinkov08,haug10,daou10,cyr-choiniere15,sato17,nakata18,auvray19,yamase21}; 
 see also \fig{dPIani}. 

On the electron-doped side, the magnetic phase may extend to a wide doping region, which could hide 
all charge ordering tendencies or make only $d$bond, $s$bond, and unibond relevant to the reality. 
As we shall show below, the $d$bond ordering tendency 
is consistent with the RXS \cite{da-silva-neto15,da-silva-neto16} and RIXS \cite{da-silva-neto18} measurements.

\subsection{Bond-charge excitations} 
The charge excitation spectrum is computed from \eq{dyson}. To extract the bond-charge component,  
we need to project $D_{a b}^{-1}$ onto the corresponding eigenvector. 
We obtain the $d$bond, $s$bond and flux order susceptibilities as follows: 
$\chi_{d{\rm bond}}^{-1} (\vq,\omega)=(1/N)(\delta/2)^{-2}(D_{33}^{-1}+D_{44}^{-1}-2 D_{34}^{-1})/2$, 
$\chi_{s{\rm bond}}^{-1}(\vq, \omega)=(1/N)(\delta/2)^{-2}(D_{33}^{-1}+D_{44}^{-1}+2 D_{34}^{-1})/2$, 
and $\chi_{\rm flux}^{-1}(\vq, \omega)=(1/N)(\delta/2)^{-2}(D_{55}^{-1}+D_{66}^{-1} -2 D_{56}^{-1})/2$. 
The difference between $\chi_{d{\rm bond}}$ and $\chi_{s{\rm bond}}$ lies in the sign 
in front of $D_{34}^{-1}$ and 
both quantities become identical for $\vq=(\pi, q_y)$ and $(q_x,\pi)$. 
Since \fig{bondphase} cannot capture the charge order around $\vq=(0.6\pi,0)$ 
reported in RIXS for the hole-doped cuprates \cite{ghiringhelli12,chang12,achkar12}, 
we here present theoretical results for the electron-doped cuprates.  
Theoretically the electron-doped cuprates are addressed by taking a positive value of 
$t'/t$ in the $t$-$J$ model \cite{tohyama94,gooding94}.

\begin{figure}[th]
\centering
\includegraphics[width=8cm]{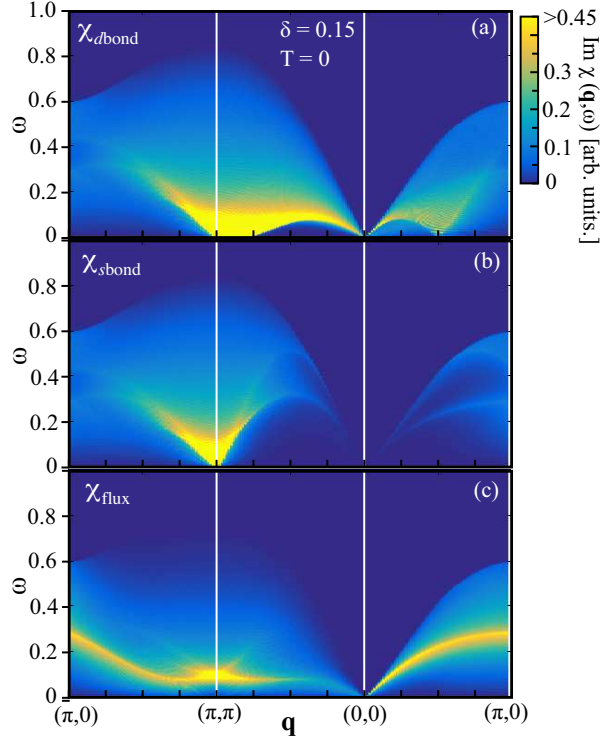}
\caption{(color online) Typical bond-charge excitations for (a) $d$bond, (b) $s$bond, and (c) flux order 
calculated in the large-$N$ theory of the $t$-$J$ model along the symmetry axes 
for doping $\delta=0.15$ at zero temperature. 
The energy unit is taken as $t$. 
Adapted from Ref.~\citen{bejas17}, where $J/t=0.3$ and $t'/t=0.3$ were employed. 
}
\label{bondspectrum}
\end{figure}

Bond-charge excitation spectra of $\chi_{d{\rm bond}}(\vq,\omega)$, 
$\chi_{s{\rm bond}}(\vq,\omega)$, and  $\chi_{\rm flux}(\vq,\omega)$ are shown in 
\fig{bondspectrum} along the symmetry axes. 
$\chi_{d{\rm bond}}$ exhibits large spectral weight at low energy around $\vq=0.8(\pi,\pi)$. 
This spectral weight is associated with the leading 
$d$bond instability at $\delta_c = 0.133$ in \fig{bondphase}. 
Along the direction $(0,0)$-$(\pi,0)$ the spectrum has rather high intensity and its energy goes down 
toward the momentum $\vq=(0.5\pi,0)$. 

$\chi_{s{\rm bond}}$ shows a low-energy dispersion around $\vq=(\pi,\pi)$, which is 
related to the proximity to the corresponding instability at $\delta_c = 0.114$ (see \fig{bondphase}). 
Its spectral weight disperses upwards forming a V-shape and loses intensity with increasing 
$\omega$. In contrast to the case of $\chi_{d{\rm bond}}$, 
there is no ordering tendency along the $(0,0)$-$(\pi,0)$ direction. 
 
$\chi_{\rm flux}$ exhibits large spectral weight at $\vq =(\pi,\pi)$ around $\omega = 0.1t$. 
This energy is reduced to zero with decreasing doping towards 
$\delta_c=0.093$, where the flux phase instability occurs (see \fig{bondphase}). 
Interestingly, there is a clear gapless dispersion along the $(0,0)$-$(\pi,0)$ direction and it extends up 
to $\omega \approx 0.3t$ at $\vq=(\pi,0)$. 
This is not a collective mode, but a peak structure of individual excitations. 
This dispersive feature continues to the $(\pi,0)$-$(\pi,\pi)$ direction and 
merges into the large spectral weight at $\omega \approx  0.1t$ and $\vq=(\pi,\pi)$.

The important insight obtained here is that the low-energy charge excitations are not dominated by 
a certain bond-charge order, but by various types  of bond-charge orders especially around $\vq=(\pi, \pi)$. 
RIXS measurements in such a region have not been performed. What was extensively studied is a region 
along the $(0,0)$-$(\pi,0)$ direction by RXS \cite{da-silva-neto15,da-silva-neto16}. 
While a signature of flux phase has not been reported, 
RXS \cite{da-silva-neto15,da-silva-neto16} and RIXS \cite{da-silva-neto18} experiments 
reported the charge ordering tendency around $\vq = (0.5\pi,0)$ 
as seen in \fig{bondspectrum}(a). 
Detailed theoretical studies \cite{yamase15b,yamase19} showed that the temperature and doping dependences of the 
spectrum are well captured in terms of the $d$bond charge excitations. 

\section{Plasmons}
Plasmons are a well established concept in metals in the presence of the long-range Coulomb interactions \cite{ashcroft} 
and were discussed theoretically also in cuprates \cite{ruvalds87,griffin88,prelovsek99,markiewicz08,vanloon14} 
including a possible superconducting mechanism \cite{kresin88,bill03} and coupling to phonons \cite{falter94,bauer09}. 
In cuprates, plasmons and their dispersion were reported around 1 eV in early studies 
by electron energy-loss spectroscopy (EELS) \cite{nuecker89,romberg90}. 
On the other hand, recent RIXS for Nd$_{2-x}$Ce$_x$CuO$_4$ (NCCO) revealed a dispersive signal, 
which has energy 0.3 eV around $\vq=(0,0)$ and increases to 1 eV around $\vq=(0.3\pi, 0)$ (Refs.~\citen{ishii05,wslee14,ishii14}). 
The origin of this signal is controversial. 
One scenario proposed intraband particle-hole excitations with strong incoherent character, 
not plasmons \cite{ishii05,ishii14,ishii17}. 
Another one is to invoke a new collective mode near a quantum phase transition 
specific to electron-doped cuprates and hence no presence of the corresponding signal in 
hole-doped cuprates \cite{wslee14,dellea17}.  
Moreover, charge excitations around $\vq=(0,0)$ were reported as  featureless and momentum-independent ones 
in the recent momentum-resolved EELS \cite{mitrano18,husain19}, which is in sharp contrast to early EELS studies  \cite{nuecker89,romberg90} and recent RIXS data \cite{ishii05,ishii14,wslee14,ishii17,dellea17,hepting18,jlin20,nag20,singh20}; 
see also Ref.~\citen{fink21}.

By taking the layered structure in cuprates and the long-range Coulomb interaction into account, 
the large-$N$ theory of the layered $t$-$J$ model on a square lattice 
turned out to explain the charge excitations observed around $\vq =(0,0)$ 
in terms of acoustic-like plasmons nearly quantitatively \cite{greco16,greco19,greco20,nag20}. 
The plasmons are generated by on-site charge fluctuations, not by bond-charge ones. 
The on-site charge excitations are described by the usual density-density correlations and given in the large-$N$ theory by 
\be
\chi_{c}(\vq, q_z,\omega) = N \left( \frac{\delta}{2} \right)^2 D_{11}(\vq, q_z,\omega) 
\ee
after summing all contributions up to $O(1/N)$. 
 Note that the large-$N$ theory can handle both on-site charge and bond-charge excitations 
 on an equal footing by choosing appropriate components $a$ and $b$ in \eq{dyson} (Ref.~\citen{bejas17}). 
 Results mentioned in Sec.~3, for which $a,b=3-6$ are taken, 
 do not change much even if the interlayer hopping and the long-range Coulomb interaction 
 are included in the $t$-$J$ model. 

\begin{figure}[th]
\centering
\includegraphics[width=14cm]{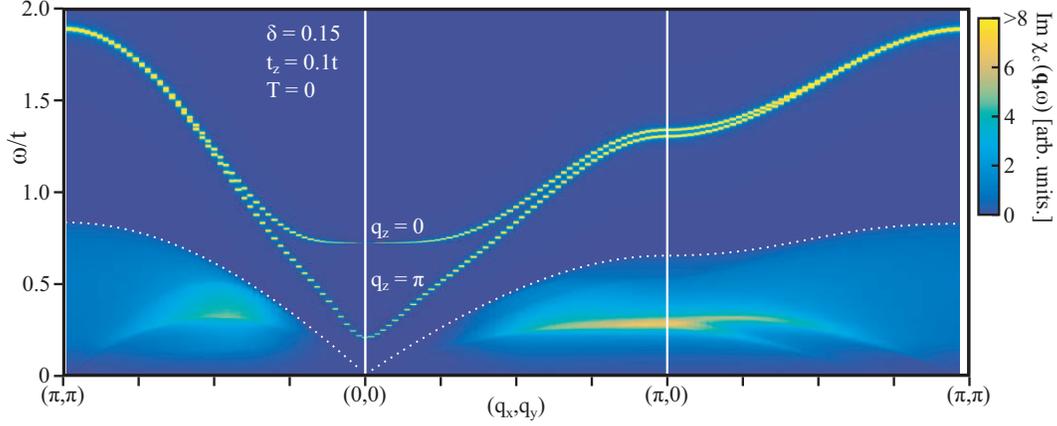}
\caption{(color online) Typical charge excitation spectrum from the usual on-site charge fluctuations 
along the symmetry axes for $q_z=0$ and $\pi$ 
computed in the large-$N$ theory of the layered $t$-$J$ model with the long-range Coulomb interaction 
for $\delta=0.15$ at zero temperature; the interlayer hopping integral 
is taken as $t_z=0.1t$. The dotted line denotes the upper boundary of a particle-hole continuum 
for $q_z=0$. Adapted from Ref.~\citen{greco16}, where $J/t=0.3$ and $t'/t=0.3$ were employed. 
}
\label{plasmon-map}
\end{figure}

Figure~\ref{plasmon-map} shows a map of the spectral weight Im$\chi_{c}(\vq,q_z,\omega)$ in the plane of 
excitation energy $\omega$ and in-plane momentum $\vq$ along 
the symmetry axes of $(\pi,\pi)$-$(0,0)$-$(\pi,0)$-$(\pi,\pi)$. 
Below $\omega \approx 0.8t$, there is a particle-hole continuum coming from 
individual charge excitations. 
The continuum does not depend much on $q_z$ and the result for $q_z=0$ is presented. 
We find no strong spectral weight near zero energy, implying that there is no (on-site) charge order 
tendency, which contrasts with the case of bond-charge orders (see \fig{bondspectrum}).  
In a high energy region, there is a sharp peak for $q_z=0$. This is the well-known 
optical plasmon and reproduces the early EELS data \cite{nuecker89,romberg90}. 

The plasmon dispersion changes drastically around $\vq=(0,0)$ once $q_{z}$ becomes finite \cite{grecu73,fetter74,grecu75}. 
As a representative, we plot the plasmon dispersion for $q_{z}=\pi$ in \fig{plasmon-map}. 
While the plasmon dispersion remains essentially the same as that for $q_z=0$ far away from $\vq=(0,0)$, 
the plasmon energy softens substantially near $\vq=(0,0)$ and exhibits a strong dispersion 
there, in sharp contrast to that for $q_z=0$. 

It should be noted that the plasmon energy has a gap at $\vq=(0,0)$ for a finite $q_z$. 
This gap comes from the presence of the finite interlayer hopping integral $t_z$ (Ref.~\citen{grecu73,grecu75,fertig91,falter94,greco16}), 
which was missed in many theoretical 
studies \cite{fetter74,ruvalds87,griffin88,kresin88,prelovsek99,bill03,markiewicz08,vanloon14}.   
The magnitude of the gap is proportional to $t_z$ in a small $t_z$ region 
and vanishes at $t_z=0$. In this sense, we call the plasmons for a finite $q_z$ 
as acoustic-{\it like} plasmons. The $t_z$ dependence of the optical plasmon is almost negligible \cite{greco16}.

\begin{figure}[th]
\centering
\includegraphics[width=7cm]{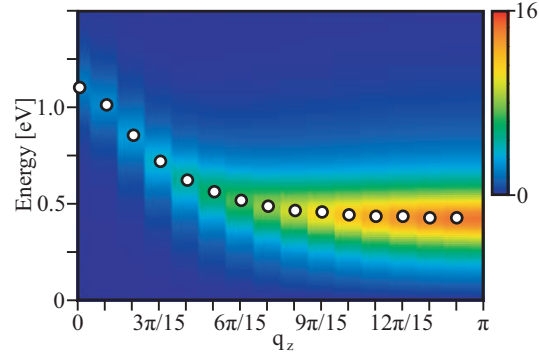}
\caption{(color online) Intensity map of plasmons as a function of $q_z$ at $\vq=0.05(\pi,\pi)$ 
computed in the large-$N$ theory of the layered $t$-$J$ model with the long-range 
Coulomb interaction. The open circles denote the peak position at each $q_z$.   
The energy is obtained from $t/2=500 \; {\rm meV}$. 
Adapted from Ref.~\citen{greco19}, where $J/t=0.3$, $t'/t=0.3$, and $t_z/t=0.1$ were employed. 
}
\label{plasmon-qz}
\end{figure}

A characteristic feature of plasmons lies in the $q_z$ dependence of the plasmon energy. 
Figure~\ref{plasmon-qz} shows a map of the spectral weight of plasmons 
in the plane of $q_z$ and $\omega$ at a small $\vq$.  
The plasmon energy rapidly decreases with increasing $q_z$ and stays almost constant in $q_z > \pi/3$; 
this rapid change is pronounced more when a smaller $\vq$ is chosen. 
The plasmon intensity, on the other hand,  increases with increasing $q_z$, following almost 
a $q_z^{2}$ dependence at small $q_z$.

\begin{figure}[b]
\centering
\includegraphics[width=7cm]{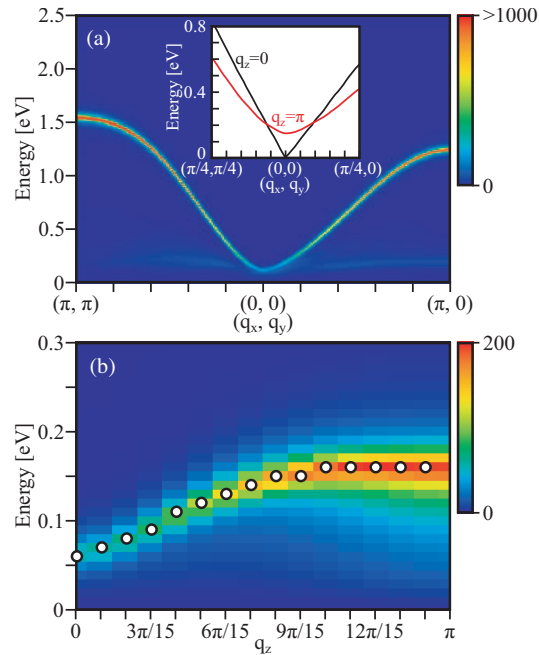}
\caption{ (color online) (a) Intensity map of charge excitations without the long-range Coulomb interaction 
in the large-$N$ theory of the layered $t$-$J$ model for $q_z=\pi$. The sharp spectrum describes 
a zero-sound mode. Inset is the zero-sound dispersion for $q_z=0$ and $\pi$ around $\vq=(0,0)$. 
(b) Intensity map of the zero-sound mode as a function of $q_z$ at a small $\vq=0.02(\pi,\pi)$. 
The open circles denote the peak position at each $q_z$.   
From Ref.~\citen{greco19}, where $J/t=0.3$, $t'/t=0.3$, and $t_z/t=0.1$ were employed.
}
\label{zero-sound}
\end{figure}

What happens if the long-range Coulomb interaction is replaced by a short-range one, 
which is more conventional in research of cuprates? 
In this case,  instead of plasmons, the zero-sound mode is realized and 
its dispersion [\fig{zero-sound}(a)] becomes similar to \fig{plasmon-map} around $\vq =(0,0)$.  
A crucial difference appears in the $q_z$ dependence. 
As shown in \fig{zero-sound}(b), 
the zero-sound mode energy {\it increases} with increasing $q_z$ at a small $\vq$, which 
is qualitatively different from the plasmon case shown in \fig{plasmon-qz}.  
This is because the zero-sound mode becomes gapless at $\vq=(0,0)$ and $q_z=0$ [inset in \fig{zero-sound}(a)], 
whereas the plasmon has a large gap as the optical plasmon (\fig{plasmon-map}).

It was shown \cite{greco19} that the acoustic-like plasmon dispersion obtained theoretically 
well reproduces the peak position of the charge excitations observed 
in electron-doped cuprates \cite{ishii05,ishii14,wslee14} with doping $\delta=0.15$, 
and for hole-doped cuprates \cite{ishii17} with $\delta=0.125$ and $0.25$. 
Moreover, as seen in \fig{plasmon-qz}, a crucial feature of plasmons is recognized in their characteristic $q_z$ dependence. 
This feature was confirmed not only in electron-doped cuprates \cite{hepting18} but also 
in hole-doped cuprates \cite{nag20}. 
Hence the charge excitations around $\vq=(0,0)$ can be summarized as follows. 
They originate from the acoustic-like plasmons 
for a finite $q_z$ and become the usual optical plasmon for $q_z=0$. 
The former explains many RIXS data \cite{ishii05,ishii14,wslee14,ishii17,dellea17,hepting18,jlin20,nag20,singh20} and 
the latter reproduces the early data of plasmons \cite{nuecker89,romberg90}. 
The acoustic-like plasmons have a gap at $\vq=(0,0)$, which has not been confirmed in experiments \cite{nag20}.

\section{Summary and outlook} 
Given that high-temperature cuprate superconductors are realized by carrier doping into a Mott insulator, 
the understanding of the charge dynamics is definitely indispensable to the cuprate physics. 
In this article, we have focused on the electronic nematic order, bond-charge orders, and plasmons. 

A crucial insight obtained in theory is that cuprates can be close to the electronic nematic instability 
\cite{yamase00a,yamase00b,yamase08}. 
The nematic order is driven by the $J$-term and the Coulomb repulsion on a square lattice and tends to be enhanced 
in the underdoped region \cite{yamase00b,okamoto10,su11,okamoto12} 
as well as around van Hove filling \cite{yamase00b,metzner00} 
in the hole-doped case; the nematic tendency is weak in the electron-doped case \cite{bejas14}. 
The nematic order is one of competing orders and can be preempted by 
antiferromagnetism and superconductivity \cite{yamase00b,grote02,honerkamp02,kampf03,edegger06,husemann12}. 
Nonetheless, even in such a case, the system can retain a large susceptibility 
to an external $xy$ anisotropy caused by, for example, crystal anisotropy, uniaxial pressure, and 
external strain \cite{yamase00b}. As a result, the system can exhibit a big $xy$ anisotropy in spite of 
a small external anisotropy \cite{yamase00a,yamase00b,yamase01,miyanaga06,yamase06,yamase07,yamase09,okamoto10,su11,okamoto12}. 
This physics traces back to a theoretical proposal in 2000 \cite{yamase00a,yamase00b}  
and is now frequently discussed as evidence of the underlying nematic order \cite{hinkov04,hinkov07,hinkov08,haug10,daou10,cyr-choiniere15,sato17,nakata18,yamase21}. 
The nematic fluctuations generate a pseudogap in the 
one-particle spectral function in a perturbative calculation to first order \cite{yamase12}. 
However, more elaborate calculations found no pseudogap \cite{yamase12}. 
Even in this case, the nematic fluctuations 
yield a large momentum dependence of the spectral weight with $d$-wave symmetry 
along the Fermi surface, leading to a Fermi-arc-like feature \cite{okamoto10,yamase12,okamoto12}. 
Given that actual materials frequently contain a small $xy$ anisotropy, the phase diagram shown in \fig{griffiths} 
may serve for a basis to discuss a global understanding \cite{yamase15}. 

Theoretical predictions about the nematic physics are supported in various experiments such as 
inelastic neutron scattering \cite{hinkov04,hinkov07,hinkov08,haug10}, ARPES \cite{nakata18}, 
Compton scattering \cite{yamase21}, 
Raman scattering \cite{auvray19}, 
measurements of Nernst coefficients \cite{daou10,cyr-choiniere15}  
and magnetic torque \cite{sato17} in hole-doped cuprates. 
Nonetheless it remains to be studied how the nematic instability and nematic fluctuations 
can be connected with the pseudogap.  
The nematic physics is also discussed in iron-based superconductors \cite{fernandes14}. 
The origin of the nematicity, however, may not lie in the $d$PI \cite{yamase00a,yamase00b,metzner00} 
nor charge stripes \cite{kivelson98}, 
but the orbital \cite{krueger09,wclee09,raghu09,cclee09,lv09} or spin nematicity \cite{blume69,andreev84,fang08,xu08}. 
This topic was not covered in this article.

Bond-charge orders range from the so-called flux phase \cite{affleck88,marston89} 
to  $s$-wave, $d$-wave, and unidirectional orders \cite{bejas12,bejas14}.  
These different charge-order tendencies are driven by the spin-spin interaction such as the $J$-term 
and can be handled on an equal footing in a large-$N$ theory of the $t$-$J$ model 
on a square lattice. 
In the hole-doped region, a large number of theoretical studies  \cite{efetov13,sachdev13,allais14,meier14,wang14,atkinson15,yamakawa15,mishra15,atkinson16,zeyher18}  
including the large-$N$ theory \cite{bejas12} were performed, but the consensus has not been obtained on the 
understanding of the RIXS data \cite{ghiringhelli12,chang12,achkar12}. 
Given the fact that 
the charge ordering tendency was observed inside the pseudogap phase \cite{keimer15}, but 
calculations did not handle the pseudogap appropriately, 
the pseudogap seems to play an important role to stabilize the charge orders. 
It is an open issue to pin down the mechanism of the charge order in hole-doped cuprates. 
On the other hand, in the electron-doped case, where the effect of the pseudogap is weak or absent \cite{armitage10}, 
the large-$N$ theory can capture the RXS \cite{da-silva-neto15,da-silva-neto16} and 
RIXS \cite{da-silva-neto18} data very well. 
It is $d$-wave bond-charge order that 
explains the charge peak observed along the $(1,0)$ direction  for various doping rates \cite{yamase15b,yamase19}. 
The large-$N$ theory also predicts large spectral weight of various bond-charge orders 
around $\vq = (\pi, \pi)$ (Refs.~\citen{yamase15b,bejas17}), 
which has not been tested in RXS nor RIXS. 

Plasmons originate from the usual on-site charge excitations in the presence of the long-range Coulomb interaction \cite{ashcroft}, 
not from the bond-charge excitations. Moreover the layered structure common to high-$T_c$ cuprate superconductors 
is needed to be considered \cite{greco16} beyond the widely studied two-dimensional models on a square lattice. 
The well-known optical plasmon \cite{nuecker89,romberg90} is obtained for $q_z=0$ and the acoustic-like plasmons 
recently reported by RIXS \cite{hepting18,jlin20,nag20,singh20} 
correspond to a finite $q_z$ with a V-shape dispersion around $\vq=(0,0)$ (Refs.~\citen{greco16,bejas17,greco19,greco20}). 
A crucial aspect is the characteristic $q_z$ dependence, which serves to identify the origin of the charge excitations \cite{greco19}. 
The plasmon scenario can explain the charge excitation spectra observed by RIXS 
for both hole- (Refs.~\citen{greco19,nag20}) and 
electron-doped \cite{greco19,greco20} cuprates almost quantitatively. 
Plasmons are the collective on-site charge excitation modes and are realized above the continuum. 
The continuum spectrum, on the other hand, does not exhibit a strong peak structure at a certain momentum 
and there is no tendency of the usual charge-density-wave instability \cite{greco16}. 

While the subjects we have discussed are limited, 
we hope that the present article serves for a sound basis toward 
further experimental and theoretical studies on the origin of the pseudogap and ultimately the high-$T_c$  mechanism. 

\hspace{5cm} 
\begin{acknowledgment}
The theoretical insights into the electronic nematic order are owed to collaboration with 
P. Jakubczyk, H. Kohno, W. Metzner, A. Miyanaga,  V. Oganesyan, and R. Zeyher. 
Theoretical studies of bond-charge orders, their fluctuations, and plasmons are 
based on collaboration with M. Bejas and A. Greco. 
The author also thanks O. K. Andersen, A. V. Chubukov, A. Eberlein, A. A. Katanin, G. Khaliullin, K. Kuboki, A. M. Oles, 
S. Sachdev, and T. Tohyama for stimulating theoretical discussions about the cuprate physics, and 
A. Fujimori, M. Fujita, M. Hepting, V. Hinkov, A. Ino, K. Ishii, B. Keimer, M. Le Tacon, A. P. Mackenzie, 
M. Minola, H. Mukuda, A. Nag,  Y. Sakurai, S. Wakimoto, K. Yamada, T. Yoshida, and K.-J. Zhou 
for fruitful discussions from an experimental point of view. 
The author is also indebted to the warm hospitality of Max-Planck-Institute for Solid State Research, without which he could not conduct his theoretical works fruitfully. 
Finally the author expresses his gratitude to M. Bejas, A. Greco, and W. Metzner for valuable comments on the present manuscript. 
This work was supported by JSPS KAKENHI Grants No.~JP20H01856. 
\end{acknowledgment}

\bibliographystyle{jpsj}
\bibliography{main} 

\begin{thebibliography}{100}
\bibitem{misc-FeSCnematic}
In the iron-based superconductors the nematic order can be regarded as a
  vestigial spin-density-wave (SDW) in the sense that the SDW is stabilized
  after the nematic instability. However, the nematic order in the iron-based
  superconductors can be understood also by a kind of orbital order, which is
  called as orbital nematic order, without invoking a vestigial order.
\end{thebibliography}

\end{document}